\documentclass[5p]{elsarticle}

\usepackage{amssymb}

\journal{TIPP09 Proceedings in NIMA}

\begin{document}

\begin{frontmatter}

\title{Results from the commissioning of the ATLAS Pixel detector}

%
\author[First]{Jedrzej Biesiada\corref{cor1}}
\ead{jbiesiada@lbl.gov}
\author[]{on behalf of the ATLAS collaboration}
\cortext[cor1]{Corresponding author. Tel.: +41-76-487-4549; fax: +1-510-486-5101.} 

%
%

\address[First]     {Lawrence Berkeley National Laboratory, Physics Division, MS50B-6222, 1 Cyclotron Road, Berkeley, CA 94720, USA}

%
%
%
%
%
\begin{abstract}

The ATLAS Pixel detector is a high-resolution, low-noise silicon-based
device designed to provide tracking and vertexing information within a
distance of 12~cm from the LHC beam axis.  It consists of
approximately 80 million pixel channels with radiation-hard front-end
electronics connected through optical fibers to a custom-controlled
DAQ system away from the detector.  Following the successful
installation of the detector in June 2007, an intense commissioning
period was conducted in the year 2008 and more than 400,000 cosmic-ray
tracks were recorded in conjunction with other ATLAS sub-detectors.
By the end of the year, $96\%$ of the detector was tuned, calibrated,
and taking data at $99.8\%$ tracking hit efficiency and with noise
occupancy at the $10^{-10}$ level.  We present here the results of the
commissioning, calibration, and data-taking as well as the outlook for
future performance with LHC collision-based data.

\end{abstract}

%
%
%
%
%
%
\begin{keyword}

Silicon pixel detector \sep
ATLAS detector commissioning



\end{keyword}

\end{frontmatter}


%
%
%
%
%
%

\section{Introduction}

The ATLAS detector~\cite{ATLAS} is one of two general-purpose
detectors that will study fundamental high-energy physics in
proton-proton collisions produced by the Large Hadron Collider (LHC).
The Inner Detector tracker is the sub-detector closest to the
beam-pipe; it consists of three components: the Transition Radiation
Tracker (TRT), the Silicon Central Tracker (SCT), and the Pixel
detector.  The Pixel detector~\cite{Pixel} is the inner-most of the
three components and the most granular, as it is designed to provide
the lowest-occupancy measurements that are critical for efficient and
accurate pattern recognition and for vertex reconstruction (including
both the primary collision vertex and secondary vertices formed by
decays of long-lived particles). The detector uses silicon pixel
technology designed to withstand the high-radiation environment of the
inner regions of the tracker.\footnote{The fluence of charged
particles at a radius of 5 cm is expected to be
$3\times10^{7}$~cm$^{-2}$~sec$^{-1}$.}

\section{Overview of the Pixel detector}

\subsection{Geometrical layout}

The Pixel detector consists of a central barrel region of three
cylindrical layers at radii of 50.5, 88.5, and 122.5~mm from the beam
axis; and two end-caps, one on each side of the barrel, consisting of
three disks located at 495, 580, and 650~mm away from the nominal
interaction point along the beam axis.  The barrel and end-caps provide
three space-point tracking in the pseudorapidity region $\eta < 2.5$.
The detector is composed of 1744 modules, each of which contains
46,080 pixel channels divided among 16 front-end chips (FE), for a
total of approximately 80 million channels.

\subsection{Pixel modules}

Pixel modules consist of $n^{+}$ pixel sensors implanted on $n$-doped
bulk wafers with a $p^{+}$ backplane, designed and qualified to
withstand an integrated radiation dose of 50~MRad.  Bulk depth is
$250~\mu$m, while most pixel dimensions are
$50~\mu$m$\times400~\mu$m.\footnote{Some pixels are larger to bridge
the regions between FE chips.}  The FE chips are bump-bonded to the
pixels and utilize $0.25~\mu$m CMOS technology to perform
pre-amplification, discrimination, measurement of time over threshold
(TOT), and digitization of the sensor signals.  A
high-density-interconnection Kapton/Copper printed circuit (flex
board) is connected to the FE chips with wire bonds and contains a
module control chip (MCC) that communicates with the chips and
multiplexes FE data, as well as performing basic error detection in
the data stream.  The flex board also contains connections to the
optical readout system and the power and HV distribution, as well as
an NTC thermistor for temperature measurement.

\subsection{Readout}

On-detector optoboards convert the electrical MCC signals from six to
seven modules into optical form with VCSEL lasers and transmit them
through optical fibers to the off-detector electronics.  The
optoboards are operated at about $20^{\circ}$C to provide low
bit-error-rate data transmission in the VCSELs.  Off-detector DAQ
boards convert the signal back to electrical form, format and monitor
the data stream, and pass it on to the global ATLAS DAQ stage.  The
same path is engaged in reverse for module configuration and control
data, with VCSELs on the DAQ boards transmitting multiplexed clock and
command information through separate optical fibers.

\begin{figure}[!tb]
\begin{center}
\includegraphics*[scale=0.35]{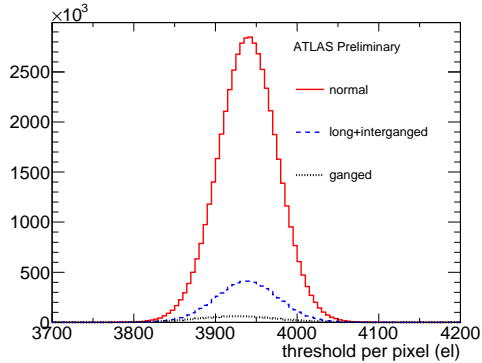}
\end{center}
\vspace{-5mm}
\caption{\label{fig:fig1}
Tuned threshold distribution for the whole detector.
}
\end{figure}

\section{Installation and calibration}

\subsection{Installation}

The Pixel package was inserted into the center of the Inner Detector
in June of 2007.  However, cable connection could not commence until
February of 2008 due to other ATLAS installation constraints and was
finished by April 2008.

The cooling system~\cite{Cooling} was commissioned loop by loop until
all 88 loops servicing the detector were operating at the end of
August 2008.  Three loops were found to have significant leaks inside
of the detector package and will be monitored as a potential source of
inefficiency in the future.  (They cannot be fixed and the
C$_{3}$F$_{8}$ coolant leaks will be studied to see if they could
cause structural defects or corrosive damage due to HF recombination
in a radiation environment after LHC turn-on.)  The evaporative
temperature in 2008 was $-10^{\circ}$C, leading to module temperatures
of roughly $-5^{\circ}$C.

\subsection{Calibration}

The exposure of the Pixel detector to high radiation doses during
operation leads to modest changes in performance of the electronics
and sensors. To retain the initial performance throughout the detector
lifetime, it is essential to regularly tune and calibrate the modules
in the detector. During 2008, we have performed a first set of
reference calibrations that will serve as our performance baseline.
The calibration consisted of tuning of the optical readout;
measurement and tuning of the signal threshold for pixel uniformity;
tuning of the TOT to set the charge response for a minimum-ionizing
particle (MIP); and TOT calibration to characterize the charge
response over the whole dynamic range of the pixels.

\subsubsection{Optical tuning}

The optical error rate is measured versus the threshold and phase of
the off-detector optical PIN receiver and versus the on-detector VCSEL
laser power. Two methods are used: a faster one, where the modules are
asked to send back 20~MHz clock and the 'one' bits are counted; and a
slower one, where the modules send back a pseudo-random bit pattern
and a bit-by-bit comparison is performed. The more accurate slower
method would only be used on problematic modules.

\subsubsection{Threshold calibration}

The pixel FE electronics have the capacity to inject charge directly
into the pixel preamplifier.  While injecting many signals, the number
of hits that are read out is counted versus the size of the injected
charge, which is scanned through the expected threshold.  The
resulting turn-on curve convoluted with roughly Gaussian noise in the
pixel is fitted to an Error function, with the fitted mean
corresponding to the threshold and the fitted width corresponding to
the noise.  The threshold is tuned pixel by pixel to the value of 4000
electrons, reducing the threshold dispersion from 100 to 40 electrons
(Fig.~\ref{fig:fig1}).  The threshold-over-noise ratio for most pixels
is around 25, resulting in low probability of noise hits.

\subsubsection{TOT tuning and calibration}
The TOT pulse width is proportional to the signal charge through a
constant-current feedback in the pixel preamplifier and is measured in
units of bunch crossings (BC), equivalent to 25~ns at the 40~MHz LHC
clock.  The TOT value can be tuned at the FE level and fine-tuned at
the pixel level.  The pixel charge response is almost linear over a
large range of deposited charge.  This behavior introduces a trade-off
between signal size and the dynamic range, since if the TOT is longer
than the ATLAS Level-1 trigger latency, the hit is lost.  The current
tuning is set to a TOT of 30~BC for a MIP charge deposition (roughly
20,000 electrons), giving a dynamic range of 8 MIPs.  The tuning has
reduced the TOT dispersion over the whole detector by $50\%$.

\subsubsection{DAQ firmware upgrade and HV leakage current}

The calibration is controlled by digital signal processors in the
DAQ boards, which run firmware binaries that are maintained and developed
within the Pixel group.  In 2008, a new version of the firmware was
introduced, incorporating a more versatile modular design and various
functional optimizations.  As a result, most calibration functions
perform two to four times faster, and new tasks have been implemented
to exploit the full calibration functionality of the FE chips.  One of
these tasks is the leakage scan, which measures the HV leakage current
pixel by pixel.  The pre-irradiation baseline shows that approximately
$99\%$ of pixels in the detector do not presently draw measurable
current.  We expect roughly 25~nA per pixel at the end of lifetime at
$-5^{\circ}$C.

\section{Taking data with cosmic rays}

When single beams were first injected into the LHC on Sept. 10, 2008,
the Pixel detector did not collect data as it was operated with HV
turned off and the FE preamplifiers disabled to protect the FE chips
from large charge depositions that are possible in unstable beam
conditions.  After the LHC incident of September 19, more than 400,000
tracks with hits in the Pixel detector were collected over several
months of data-taking with cosmic rays, interleaved with the
calibration efforts.  After masking roughly $0.01\%$ of noisy pixels
in the detector, noise occupancy at the $10^{-10}$-per-BC level was
achieved, corresponding to roughly $0.05$ noise hits per event in the
whole detector.

\subsection{Timing}

The Pixel detector was reading out a window of 8 consecutive BCs in
order to allow synchronization with the other ATLAS sub-detectors,
which was quickly achieved.  As the detector is currently timed-in to
within 3~BCs, the read-out window will be reduced to 5~BCs for initial
LHC data, and ultimately to 1-3~BCs.

\begin{figure}[!tb]
\vspace{5mm}
\begin{center}
\includegraphics*[scale=0.35]{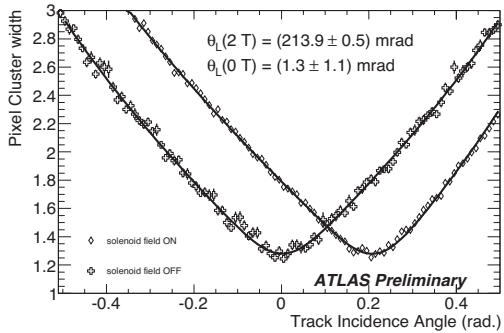}
\end{center}
\vspace{-5mm}
\caption{\label{fig:fig2} Distribution of the cluster width versus
track incidence angle for the Pixel detector for two values of the
magnetic field, showing the fitted Lorentz-angle values.  }
\end{figure}

\subsection{Calibration studies with cosmic rays}

\begin{figure}[tb]
\begin{center}
\includegraphics*[scale=0.35]{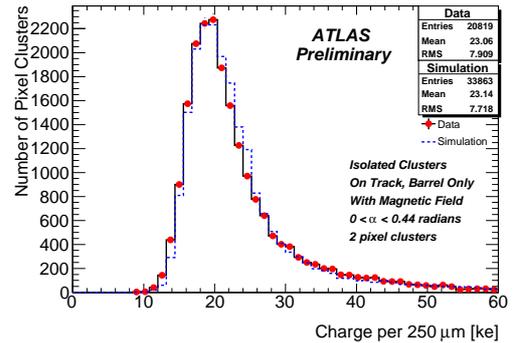}
\end{center}
\vspace{-5mm}
\caption{\label{fig:fig3} Distribution of the cluster charge
distribution in data (points and fitted solid line) and simulation
(dashed) for tracks with nearly normal incidence.}
\end{figure}

The cosmic-ray data has been extensively analyzed for additional
calibration measurements.  Fig.~\ref{fig:fig2} shows the cluster width
in the pixel detector versus incidence angle of the track for two
different values of the solenoidal magnetic field in the Inner
Detector.  The minimum of the distributions occurs at the Lorentz
angle, which reflects the deflection of the charge carriers in the
magnetic field.  This angle is consistent with zero when the magnetic
field is off and about 214~mrad for the full magnetic strength of 2~T.
This result is consistent with simulated expectations to within $5\%$.
Fig.~\ref{fig:fig3} shows that the cluster-charge distribution for
tracks with nearly normal incidence agrees very well with simulated
events as well.  Remaining disagreements are being studied to improve
modeling of the detector response.

Misalignment of the Pixel detector has been reduced to the $20~\mu$m
level in the $50~\mu$m pixel direction (azimuthal) and to $30~\mu$m in
the $400~\mu$m direction (longitudinal) after extensive alignment
efforts down to the module level in the central barrel.  End-cap
alignment will require LHC beam data.  With the most recent aligned
geometry, the efficiency of attaching a hit to a track traversing the
barrel is $99.8\%$, which is compatible with the test-beam value of
$99.9\%$~\cite{testbeam}.

\section{Conclusion and outlook}

After an intensive commissioning and cosmic-ray data-taking period
over the year 2008, $96\%$ of the ATLAS Pixel detector is tuned,
calibrated, and ready to collect data, with all but about $2\%$
expected to be recovered. The cosmic-ray data sample has
been used for extensive calibration, tuning of simulation input, and
on-going studies of detector performance.  After upgrade work on the
cooling system concludes in May 2009, the detector will be
re-calibrated in preparation for LHC collisions to be delivered by the
end of 2009.

%
%
%
%
%
%


\begin{thebibliography}{00}

%
%
\bibitem{ATLAS} The ATLAS Collaboration, G. Aad et al., The ATLAS
Experiment at the CERN Large Hadron Collider, JINST 3 (2008) S08003.
%

\bibitem{Pixel} G Aad et al., ATLAS pixel detector electronics and
sensors, JINST 3 (2008) P07007.

\bibitem{Cooling} D. Attree et al., The evaporative cooling system for
the ATLAS inner detector, JINST 3 (2008) P07003.

\bibitem{testbeam} A. Andreazza, on behalf of the ATLAS Pixel Collaboration, Nucl. Instr. Meth. A 565 (2006) 23-29.

\end{thebibliography}
\end{document}